\documentclass[runningheads]{llncs}
\usepackage{graphicx}
\usepackage{amssymb}
\usepackage{url}
\usepackage{caption}
\usepackage{subcaption}
\usepackage{listings}
\usepackage{amsmath}
\usepackage{breqn}
\usepackage{color}
\usepackage[table,xcdraw]{xcolor}
\usepackage{xcolor}
\usepackage{textcomp}
\usepackage{multirow}
\usepackage{booktabs}
\usepackage[normalem]{ulem}
\usepackage{xspace}
\useunder{\uline}{\ul}{}
 \newcolumntype{a}{ >{\columncolor[HTML]{ffccf2}} c }
 \newcolumntype{b}{ >{\columncolor[HTML]{ffffcc}} c }
 \newcolumntype{d}{ >{\columncolor[HTML]{ccddff}} c }
 \newcolumntype{e}{ >{\columncolor[HTML]{ e6fff2}} c }
  \newcolumntype{f}{ >{\columncolor[HTML]{f0f0f5}} l }
   \newcolumntype{g}{ >{\columncolor[HTML]{ e6eeff}} c }
\definecolor{Gray}{gray}{0.9}
\urldef{\mailsa}\path|{zhangxy216,meng.wang,gqi}@seu.edu.cn, saleem@informatik.uni-leipzig.de, axel.ngonga@upb.de, carter.whfcarter@gmail.com|

\makeatletter
\DeclareRobustCommand\onedot{\futurelet\@let@token\@onedot}
\def\@onedot{\ifx\@let@token.\else.\null\fi\xspace}

\def\eg{{e.g}\onedot} 
\def\ie{{i.e}\onedot} 
 
\def\etc{{etc}\onedot} 
 
\def\etal{{et al}\onedot}

\begin{document}
\captionsetup[table]{name={Tab.}}
\title{Revealing Secrets in SPARQL Session Level} 
%
%

\author{Xinyue Zhang\textsuperscript{1}
\and Meng Wang\textsuperscript{1,2,\thanks{Corresponding author}}\and Muhammad Saleem\textsuperscript{3}\and \\ Axel-Cyrille	Ngonga Ngomo\textsuperscript{4} \and Guilin Qi\textsuperscript{1,2} \and Haofen Wang\textsuperscript{5}
}

\authorrunning{Xinyue Zhang et al.}

\institute{	1. School of Computer Science and Engineering, Southeast University, Nanjing, China\\
	2. Key Laboratory of Computer Network and Information Integration (Southeast University), Ministry of Education, Nanjing, China\\
	3. AKSW, Leipzig University, Leipzig, Germany\\
	4. University of Paderborn, Paderborn, Germany\\
	5. Intelligent Big Data Visualization Lab, Tongji University, Shanghai, China\\
	\mailsa\\
}

\maketitle         
\begin{abstract}

Based on Semantic Web technologies, knowledge graphs help users to discover information of interest by using live SPARQL services. Answer-seekers often examine intermediate results iteratively and modify SPARQL queries repeatedly in a search session. In this context, understanding user behaviors is critical for effective intention prediction and query optimization. However, these behaviors have not yet been researched systematically at the SPARQL session level. 
This paper reveals secrets of session-level user search behaviors by conducting a comprehensive investigation over massive real-world SPARQL query logs.
In particular, we thoroughly assess query changes made by users w.r.t. \emph{structural} and \emph{data-driven} features of SPARQL queries. To illustrate the potentiality of our findings, 
we employ an application example of how to use our findings, which might be valuable to devise efficient SPARQL caching, auto-completion, query suggestion, approximation, and relaxation techniques in the future\footnote{Code and data are available at: \url{https://github.com/seu-kse/SparqlSession}}. 

\end{abstract}

\section{Introduction}

Semantic Web technologies enable an increasing amount of data to be published as knowledge graphs using RDF. SPARQL endpoints have emerged as useful platforms for accessing knowledge graphs via live SPARQL querying. Currently, there are billions of RDF triples available from hundreds of SPARQL endpoints~\footnote{\url{https://sparqles.ai.wu.ac.at/availability}, accessed on \today.}.
However, users often fail to express their information needs in one succinct query. This is due to their unfamiliarity with the ontology underlying the endpoints they query, or with SPARQL's syntax. This finding has been corroborated by an analysis on the LSQ dataset~\cite{2015LSQ}, where 31.70\% of the real-world queries posted to 4 different SPARQL endpoints contain parse errors and 21.42\% of the queries produce zero answers. Therefore, SPARQL queries are continuously refined to retrieve satisfying results in practice.
We can use techniques based on information about underlying data or query sequence history to assist users. The underlying data is informative and useful, but in some cases, historical queries are the main source of information that is available, \eg, where we do not have access to data.
In this paper, we provide session-level query analysis to enhance techniques based on query sequence history.

\begin{figure}[t]
\centering
  \includegraphics[width=\linewidth]{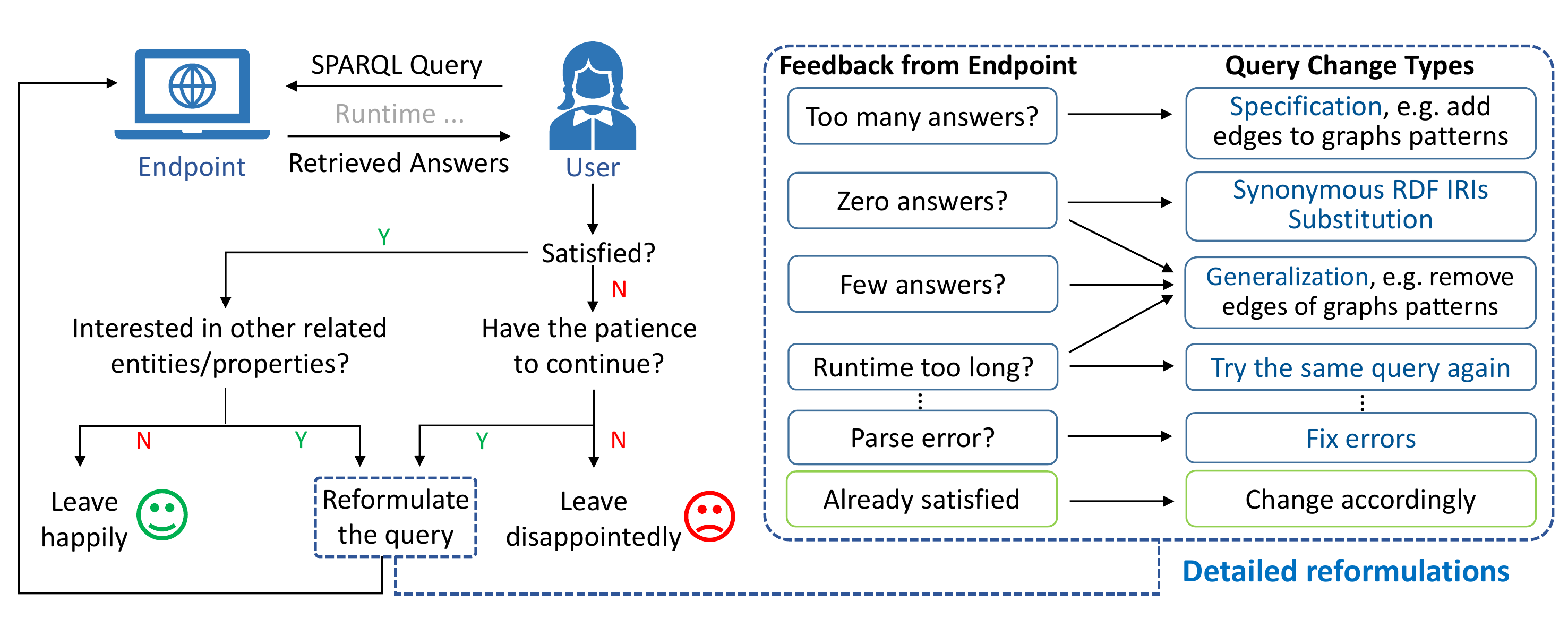}
  \caption{A typical SPARQL search session}
  \label{fig:interactive}
\end{figure}

In the field of Information Retrieval (IR), the continuous query reformulation process is called a \emph{search session} and has been well-studied to generate query suggestions and enhance user satisfaction by utilizing implicit (\eg, query changes~\cite{queryChangeModel}, clicks and dwell time in a certain website~\cite{cogSessSearch}), and explicit (\eg, relevance scores~\cite{queryChangeModel}) user feedback.
In a SPARQL \emph{search session}, feedback from users is generally only revealed in query changes, which makes it more challenging to understand drifting user intentions. Fortunately, SPARQL queries contain richer information in query \emph{change types} (Fig.~\ref{fig:interactive}) compared to the keyword queries in IR. Thus, a more detailed session-level analysis of the real-world SPARQL queries posted by users is both possible and of central importance for devising efficient caching mechanisms~\cite{2013detectSplTemp}, query relaxation~\cite{relaxation,wang2018towards}, query approximation~\cite{appro2}, query auto-completion~\cite{completion}, and query suggestion~\cite{autoSPARQL}.

Prior SPARQL query log analyses~\cite{2017pattern,2011empirical,2011realSPARQL,2015LSQ} have focused on analyzing the \emph{structural} features (e.g., usage of different SPARQL operators, triple patterns, types of joins) of queries in isolation. The potential correlations between queries in a search session have not been fully investigated. 
Similarities between queries within the same session have been reported ~\cite{bonifati2019analytical,2012MachineLoopPattern}. This property has been used in query augmentation~\cite{2013detectSplTemp} to retrieve closely related results. 
However, these works do not provide deeper insight into query changes within the search session. In addition, there has been no distinction made between robotic queries (\ie, machine-generated) and organic (\ie, human-generated) queries. Given that the distribution of queries in SPARQL endpoints is heavily dominated by robotic queries, in terms of volume and query load (over 99\% in LSQ~\cite{2015LSQ}), current studies on the similarity of queries depend heavily on robotic queries. 


In this paper, we fill the research gap discussed above via the session-level analysis of real-world organic queries collected from 10 SPARQL endpoints. Specifically, we study the evolvement of \emph{structural} and \emph{data-driven} (e.g., result set size) features in single SPARQL \emph{search sessions}. We also provide comprehensive insights regarding session-level query reformulations on SPARQL operators, triple patterns, and \texttt{FILTER} constraints. 
Furthermore, we implement an application example about the usage of our findings which might be useful to devise more efficient mechanisms for SPARQL query auto-completion, recommendations, caching, \etc.
Our contributions can be summarized as follows:
\begin{itemize}
    \item 
    We port the concept of sessions to SPARQL queries and give a specification of SPARQL \emph{search sessions}.
    \item We are the first, to the best of our knowledge, to investigate potential correlations between SPARQL queries and provide a comprehensive analysis of query reformulations in a given search session.
    \item 
    We provide an application example of how our findings can be used to illustrate the potentiality of utilizing user behaviors in a search session.
\end{itemize}

\section{Preliminaries}
\label{sec:prelim}
This section briefly introduces datasets and the pre-processing we use, followed by a formal definition of the SPARQL \emph{search session}, as well as \emph{structural} and \emph{data-driven} features of SPARQL queries.

\subsection{Datasets and Pre-processing}
\label{sec:dataset}
The difficulties of formulating a SPARQL query depend on the complexity of schema of knowledge graphs. Also, SPARQL queries that are used to query knowledge graphs from different domains have different features.
Therefore, we selected 10 LSQ datasets~\cite{2015LSQ} (version $2$, $15$ from Bio2RDF~\cite{bio2rdf} and $3$ others), containing real-world SPARQL queries collected from public SPARQL endpoints of these datasets. The selected datasets include 7/15 diverse datasets from Bio2RDF~\cite{bio2rdf} (a compendium of bioinformatics datasets in RDF), \ie, NCBI Gene (Ncbigene), National Drug Code Directory (Ndc), Orphanet, Saccharomyces Genome Database (Sgd), Side Effect Resource (Sider), Affymetrix, Gene Ontology Annotation (Goa), and the remaining 3 datasets: DBpedia~\cite{bizer2009dbpedia} (extracted from Wikipedia), SWDF~\cite{swdf} (Semantic Web Dog Food about conferences metadata), and LinkedGeoData (LGD)~\cite{linkedgeodata} (a spatial RDF dataset).



\begin{table}[h]
\centering
\caption{Statistics of SPARQL query log datasets. (The `` /\ " is used to show the number of queries (executions) \emph{excluding/including} parse errors, while colors are for different domains.) }
\label{tab:basicInfo}
\resizebox{\textwidth}{!}{%
\begin{tabular}{lccrrr}
\toprule
Dataset & \#Queries & \#Executions & Begin time & End time & \#Users \\
\midrule
\rowcolor[HTML]{ffccf2} 
LGD & 651,251/667,856 & 1,586,660/1,607,821 & 2015/11/22 & 2016/11/20 & 26,211 \\
\rowcolor[HTML]{ffffcc} 
SWDF & 520,740/521,250 & 1,415,438/1,415,993 & 2014/5/15 & 2014/11/12 & 936 \\
\rowcolor[HTML]{ccddff} 
DBpedia & 3,001,541/4,196,762 & 3,552,212/6,248,139 & 2015/10/24 & 2016/2/11 & 39,922 \\
\rowcolor[HTML]{e6fff2} 
Affymetrix & 618,796/630,499 & 1,782,776/1,818,020 & 2013/5/5 & 2015/9/18 & 1,159 \\
\rowcolor[HTML]{e6fff2} 
Goa & 630,934/638,570 & 2,345,460/2,377,718 & 2013/5/5 & 2015/9/18 & 1,190 \\
\rowcolor[HTML]{e6fff2} 
Ncbigene & 679,586/689,885 & 1,561,592/1,593,958 & 2014/5/14 & 2015/9/18 & 417 \\
\rowcolor[HTML]{e6fff2} 
Ndc & 707,579/720,838 & 2,354,808/2,411,232 & 2013/5/16 & 2015/9/18 & 1,286 \\
\rowcolor[HTML]{e6fff2} 
Sgd & 618,670/630,891 & 1,992,800/2,038,097 & 2013/5/5 & 2015/9/18 & 1,304 \\
\rowcolor[HTML]{e6fff2} 
Sider & 186,122/187,976 & 677,950/681,247 & 2015/5/31 & 2015/9/18 & 216 \\
\rowcolor[HTML]{e6fff2} 
Orphanet & 476,603/477,036 & 1,521,797/1,523,459 & 2014/6/11 & 2015/9/18 & 171 \\
\bottomrule
\end{tabular}%
}
\end{table}

\begin{figure}[!htb]
  \includegraphics[width=\linewidth]{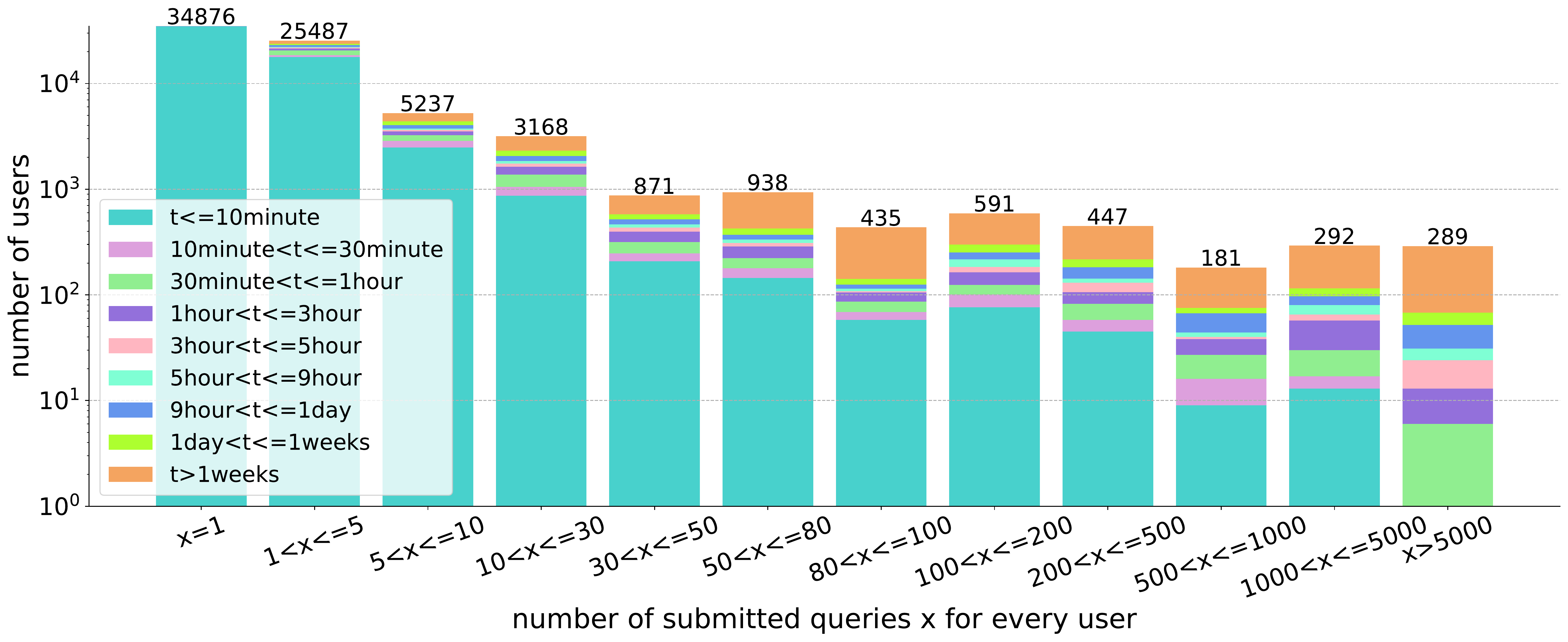}
  \caption{Distribution of the number of submitted queries $x$ and time-span $t$ of the submitted queries for each user. The X-axis indicates different intervals of submitted query numbers; The Y-axis indicates the number of users in the interval. Different colors indicate different time-spans.}
  \label{fig:combined_bar}
\end{figure}


Tab.~\ref{tab:basicInfo} gives an overview of the selected datasets in terms of the number of queries (\texttt{\#Queries}) and their total number of executions (\texttt{\#Executions}\footnote{A query can be executed multiple times on the same dataset.}) executed by different users (\texttt{\#Users}) within a time frame.
The basic distribution of the number and time-span of the submitted queries for each user is presented in Fig.~\ref{fig:combined_bar}. 
This figure shows the existence of robotic queries that are submitted in a short period. These robotic queries do not show clear trends in individual human usage~\cite{2018wikidata} but easily cause analytic biases due to their sheer size. 
Therefore, we need to remove robotic queries. 
There are generally three characteristics that can be used to recognize them: (1) special agent names (\eg Java)~\cite{2018wikidata,2012MachineLoopPattern} (2) relatively high query request frequency~\cite{2012MachineLoopPattern} (3) loop patterns existing in query sequences submitted by one user~\cite{2012MachineLoopPattern}, where the SPARQL structures remain the same in contiguous queries, while only IRIs, literals, or variables change. 
%
However, due to the privacy policy, agent names are usually unavailable in practice.
Therefore, we combine (2) and (3) to design a two-step process: (a) filtering out users who submit queries with a high-request frequency, \ie, users who submit more than $30$ times in a $30$ minutes sliding window. 
This threshold is a relatively high frequency in our dataset. We use it to compute the average frequency.
Also, to make sure this is not a rigorous cut-off rule, we supplement the second step:
(b) examining every query sequence submitted by one user. Those sequences with loop patterns are excluded. 
After robotic query removal process, there are $51,575$ ($0.64\%$) likely organic queries and $67,594$ ($0.36\%$) executions in our datasets. These executions come from $7,718$ ($10.60\%$) users, each having submitted $8.76$ queries on average.


\subsection{Definitions}
\label{subsec:def}
\begin{figure}[t]
\begin{minipage}[t]{0.49\linewidth}
\centering
\includegraphics[width=\linewidth]{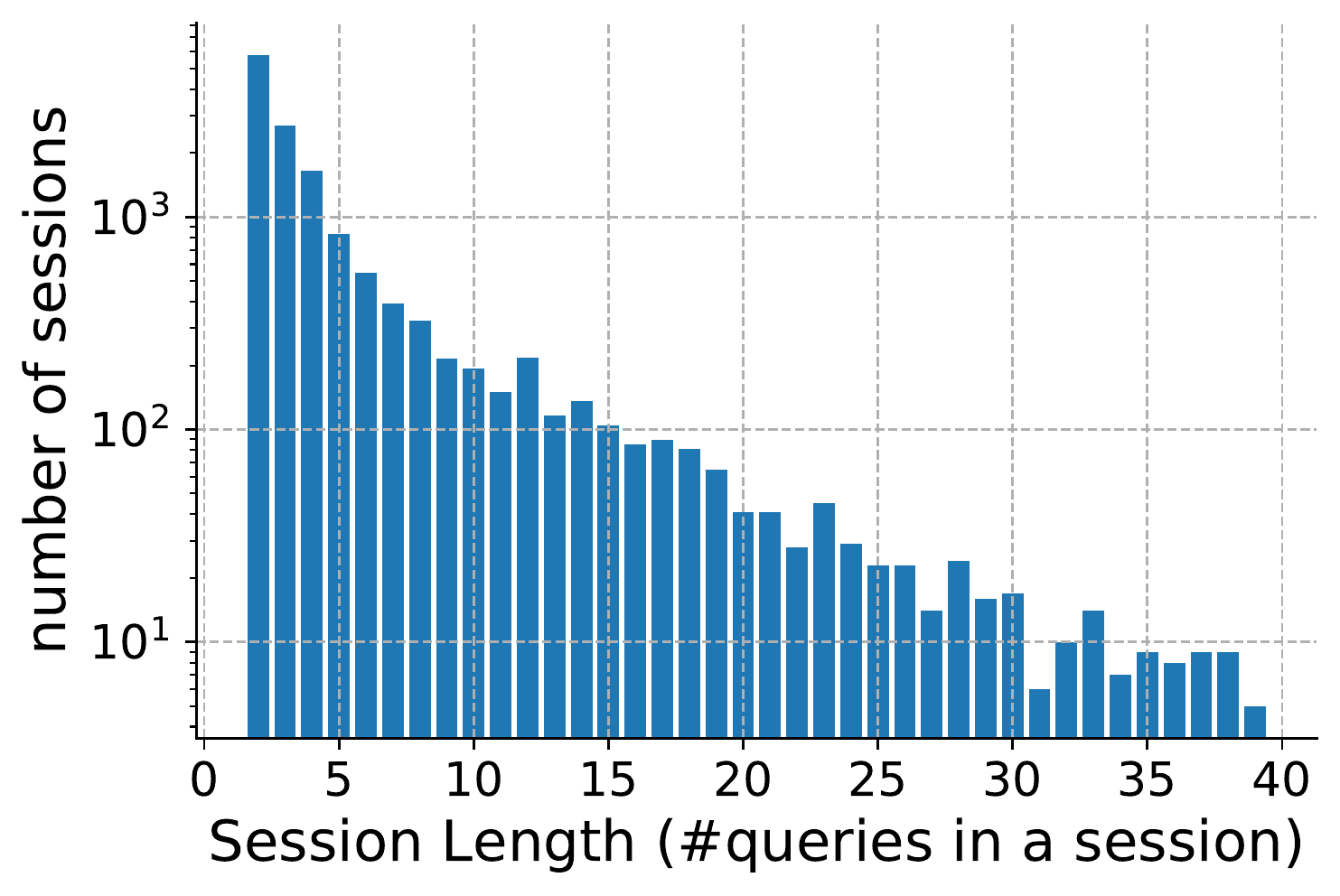}
\caption{Organic session length.}
\label{fig:sessionLen}
\end{minipage}%
\hfill
\begin{minipage}[t]{0.49\linewidth}
\centering
\includegraphics[width=\linewidth]{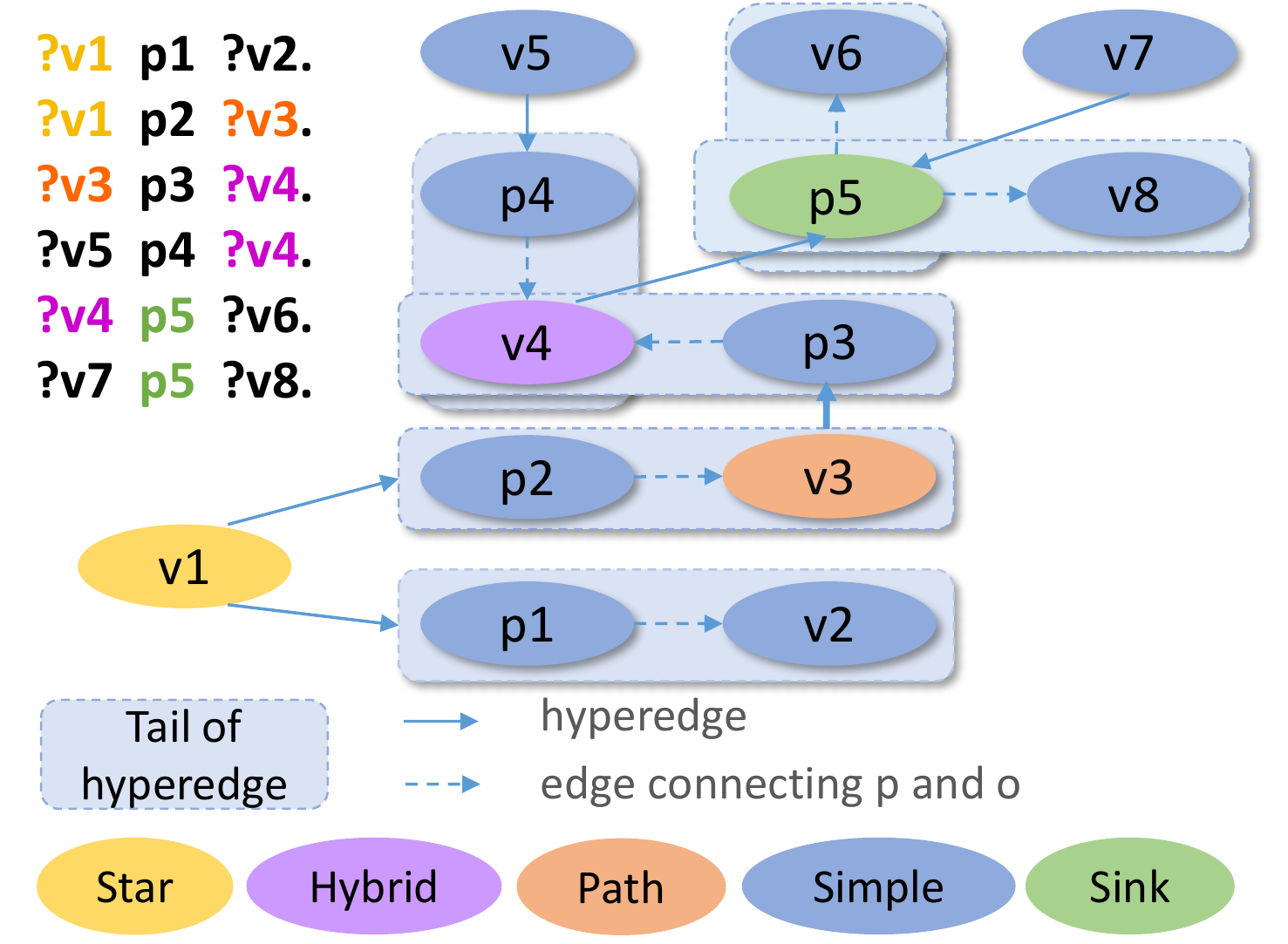}
\caption{Hypergraph of a BGP.}
\label{fig:graph_pre}
\end{minipage}%
\end{figure}

Formally, a \textbf{SPARQL search session} $s = \{Q, R, T\}$ consists of a sequence of queries $Q = \{q_1,\cdots, q_i,\cdots,q_n\}$, retrieved result sets $R = \{R_{q_1}, \cdots, R_{q_i}, \cdots, R_{q_n}\}$, and time information $T = \{T_{q_1},\cdots, T_{q_i}, \cdots, T_{q_n}\}$, where $n$ is the number of queries in the session (i.e., the session length) and $i$ indexes the queries. Each result set $R_{q_i}$ contains all the results of $q_i$, while each time information $T_{q_i}$ contains the time stamp and executing runtime of $q_i$. In practice, a SPARQL search session is recognized by three constraints\footnote{We remove the queries with parse errors and the contiguous same queries before the recognition. For example, $q_1, q_2, q_2, q_3$ is processed into $q_1, q_2, q_3$.}: queries in a sequence are (1) executed by one user, which is identified by encrypted IP addresses (2) within a time window of a fixed $time\_threshold$ (inspired by~\cite{2013detectSplTemp,2018Session}, we set $time\_threshold$ to 1 hour in this paper). (3) If we define $term(q)$ (i.e., a term set of one query) as a set which contains all the specified terms (i.e., RDF IRIs) and variables used in the query, then for any pair of contiguous queries $(q_i, q_{i+1})$ in the session, it satisfies $term(q_i)\cap term(q_{i+1}) \neq \varnothing $. 
Here, we include variables in the term set $term(q)$ because our experiments shows that users typically do not change variable names in a session: 
91\% (27\% for 1 variable, 35\% for 2 variables, and 29\% for more than 2 variables) of continuous query pairs in sessions have at least one variable name in common.
%
Please note that, (3) here is a minimum requirement for sessions, while the one user and 1-hour threshold setting can ensure the topic continuity to a large extent, which can be evaluated by the number of common variable names and high similarity score of IRI terms in Fig.~\ref{fig:kl_cos} (introduce later). 
Furthermore, although we acknowledge that there could be other ways to identify sessions, the method we present here is reasonable.
Based on these constraints of sessions, we extract $14,039$ sessions from organic queries in our dataset. The distribution of the organic session length is presented in Fig.~\ref{fig:sessionLen}.

We follow~\cite{saleem2018largerdfbench,watdiv2014,saleem2019representative,FEASIBLE2015} to define two types of SPARQL query features, \ie, \emph{structural} and \emph{data-driven} features, for the SPARQL session-level analysis. 

\noindent\textbf{Structural features:} The basic graph patterns (BGPs) in SPARQL queries organize a set of triple patterns into different types of structures. We represent each BGP of a SPARQL query as a directed hypergraph to easily compare the structural changes between different queries in a search session. The hypergraph representation~\cite{saleem2019representative} contains nodes for all three components of the triple patterns $<$\emph{s,p,o}$>$. A hyperedge $e = (s,(p,o)) \in E \subseteq V^3$ connects the head vertex $s$ and the tail hypervertex $(p, o)$, where $E$ is the set of all hyperedges and $V$ is the set of all vertices in the hypergraph.
The hypergraph of a complete SPARQL query (consider BGPs only) is the union of hypergraph representations of all BGPs in the query. An example is illustrated in Fig.~\ref{fig:graph_pre}. We define the following \emph{structural} features based on the hypergraph representation.

\begin{itemize}
\item The \textbf{triple pattern count} refers to the number of triple patterns in a BGP, which distinguishes between simple and structural complex queries.

\item The \textbf{join vertex type} is characterized by in-degrees and out-degrees for vertices in BGPs. A \emph{star} vertex only has (multiple) outgoing edges. A \emph{sink} vertex only has (multiple) incoming edges. There is only one in-degree and one out-degree for a \emph{path} vertex. The in-degree (or out-degree) of a \emph{hybrid} vertex is more than one while out-degree (or in-degree) is at least one. 

\item The \textbf{join vertex degree} indicates the summation of the in-degree and out-degree of a join vertex. For a SPARQL query, we use the minimum, mean, and maximum of join vertex degrees in the query to represent this feature.

\item The \textbf{projection variable count} is the number of selected variables that form the solution sequences in the \texttt{SELECT} query form.

\item The \textbf{IRI term set} is the collection of used IRI terms in a SPARQL query. This feature presents the information that users are interested in.
\end{itemize}

\noindent\textbf{Data-driven features:} We mainly consider the \textbf{result size}, \ie, $|R_{q_{i}}|$, the number of solutions for SPARQL queries in this paper. The change of result size (decrease or increase) generally reflects whether users want more specific or more general answers, and as a result, is an important feature to capture the drifting query intentions.



\section{Query Changes in SPARQL Search Session}
\label{sec:ana}
User search behaviors, represented by changes over queries, are the key to understand user intentions. In this section, we investigate query changes based on the aforementioned SPARQL query features from two aspects: (1) the evolvement of query changes (Sec.~\ref{subsec:evolvement}); (2) detailed reformulation strategies (Sec.~\ref{subsec:op}). 
Please note that due to the space limitation, we only provide individual dataset-level results when different datasets show very different results. More rudimentary dataset analysis are provided in~\cite{2015LSQ}.


\subsection{Evolvement of Query Changes}
\label{subsec:evolvement}
We study the query evolvement in terms of three structural aspects (\ie, graph edit distance, graph pattern similarity, and IRI term similarity), as well as one data-driven aspect, \ie, changes of result size.

\begin{figure}[t]
  \centering
  \begin{subfigure}[b]{0.49\linewidth}
    \includegraphics[width=\linewidth]{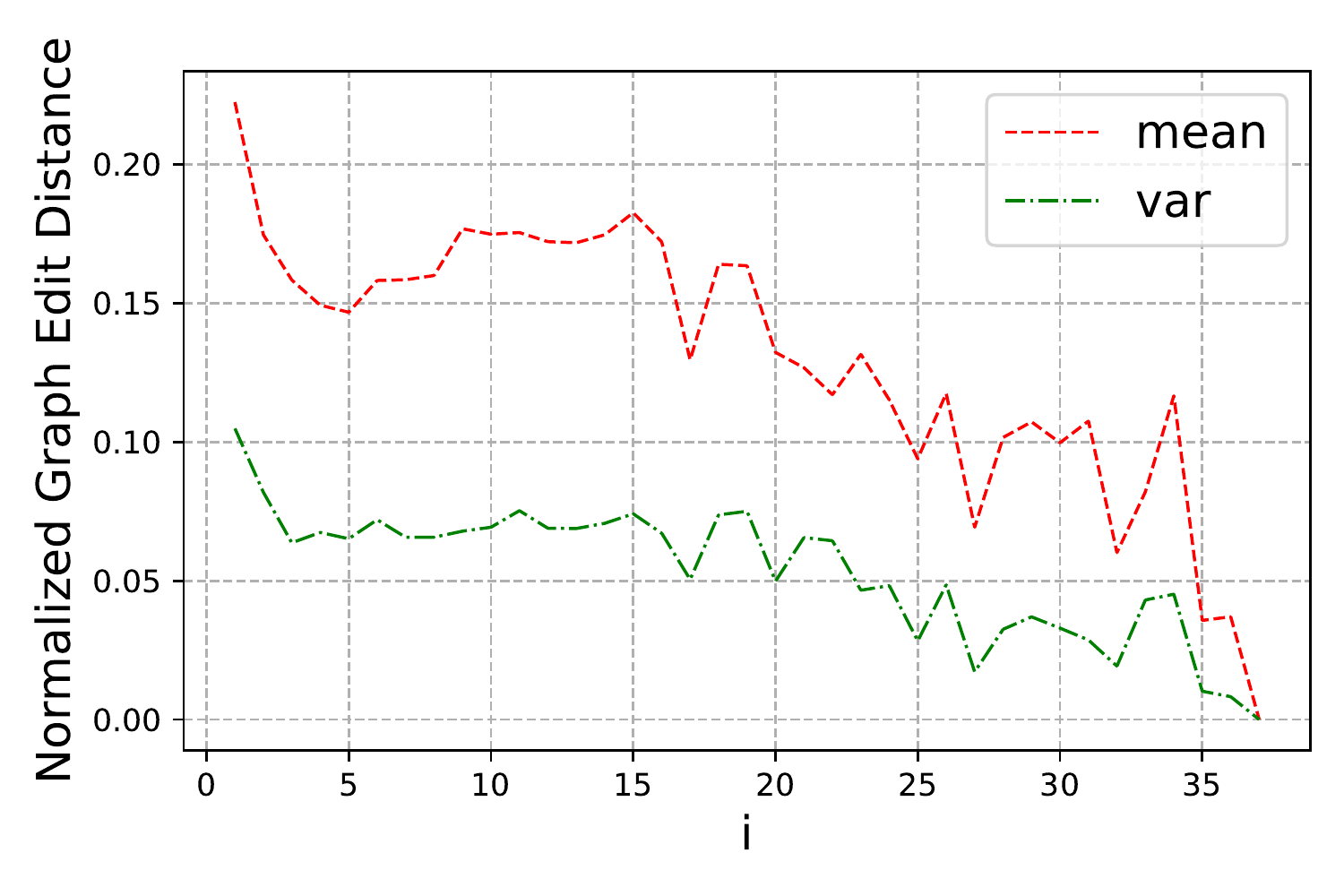}
     \caption{GED for $(q_i, q_{i+1})$}
     \label{subfig:conti}
  \end{subfigure}
  \begin{subfigure}[b]{0.49\linewidth}
    \includegraphics[width=\linewidth]{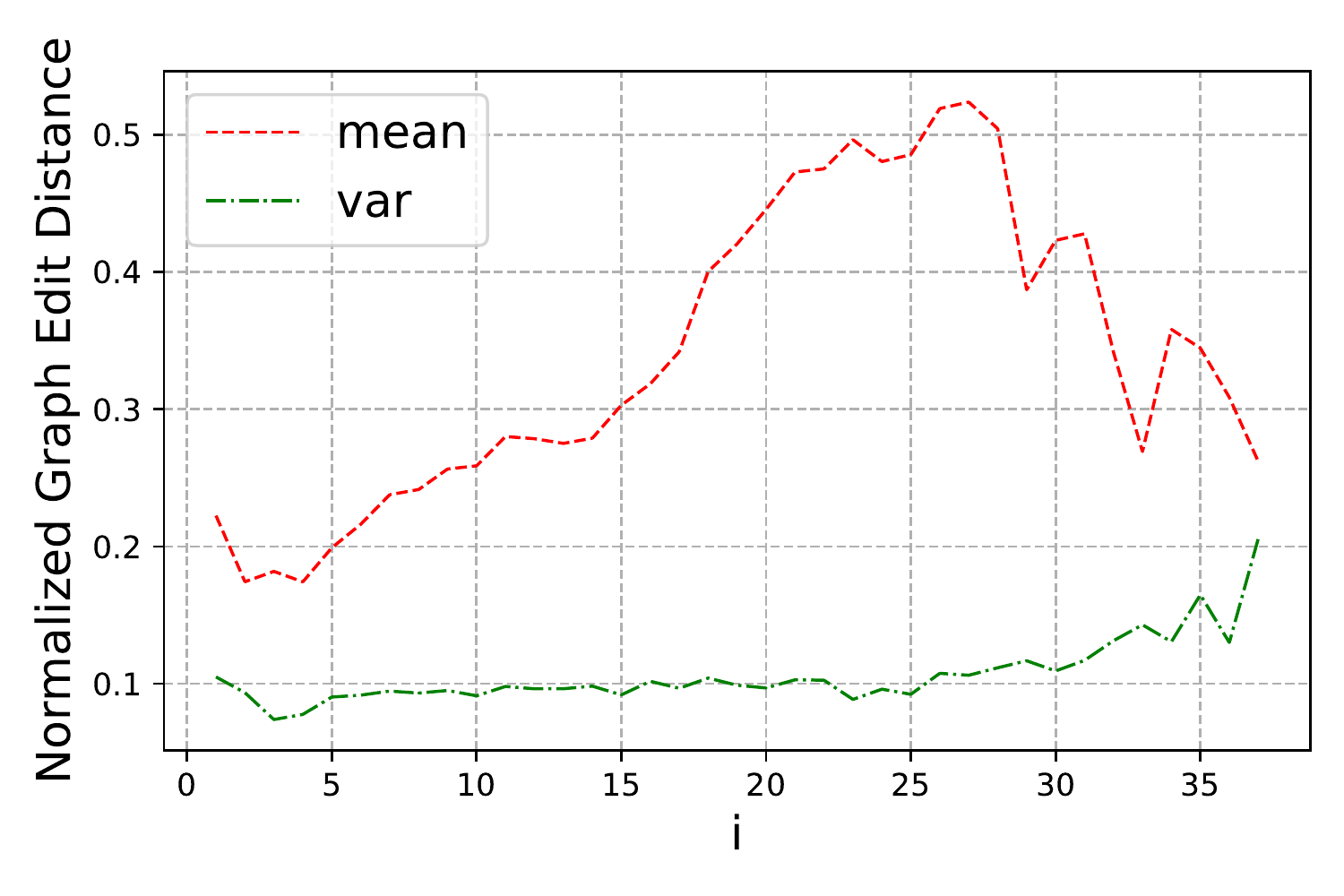}
    \caption{GED for $(q_1, q_{i+1})$}
    \label{subfig:first}
  \end{subfigure}
\caption{Evolvement of GED of $Q$ in sessions. X-axis shows the query index `` i\ ". }
\label{fig:conti_first_graph}
\end{figure}

\noindent\textbf{Graph Edit Distance (GED):} Given a query sequence $Q$ of a session $s$, we represent each BGP of queries as a directed hypergraph and utilize the normalized GED to measure differences between hypergraphs. The GED between two hypergraphs is normalized by dividing the maximum of the number of edges and vertices in two hypergraphs. On this basis, the GED between two queries is accessed by the average of GEDs between hypergraphs in different operator blocks of the two queries. A GED numeric value ranges from $0$ (no change), to $1$ (complete change), indicating the degree of changes between two queries. We conduct two types of comparisons: (1) GED between two contiguous queries, \ie, $(q_i, q_{i+1})$, (2) GED between the initial query and the other query in a given session, \ie, $(q_1,q_{i+1})$. Consider the sequence $\{q_1, q_2, q_3\}$ as an example: we calculate GED values of (1) $(q_1,q_2)$,$(q_2,q_3)$, and (2) $(q_1,q_2)$,$(q_1,q_3)$. The average and variance of GEDs (given by \emph{mean} and \emph{var} respectively) in single search sessions on our 10 datasets\footnote{We use randomly selected sessions in DBpedia because the GED computation on such large-scale data is NP-hard and time-consuming.} are presented in Fig.~\ref{fig:conti_first_graph}.


\begin{figure}[t]
  \centering
  \begin{subfigure}[b]{0.49\linewidth}
    \includegraphics[width=\linewidth]{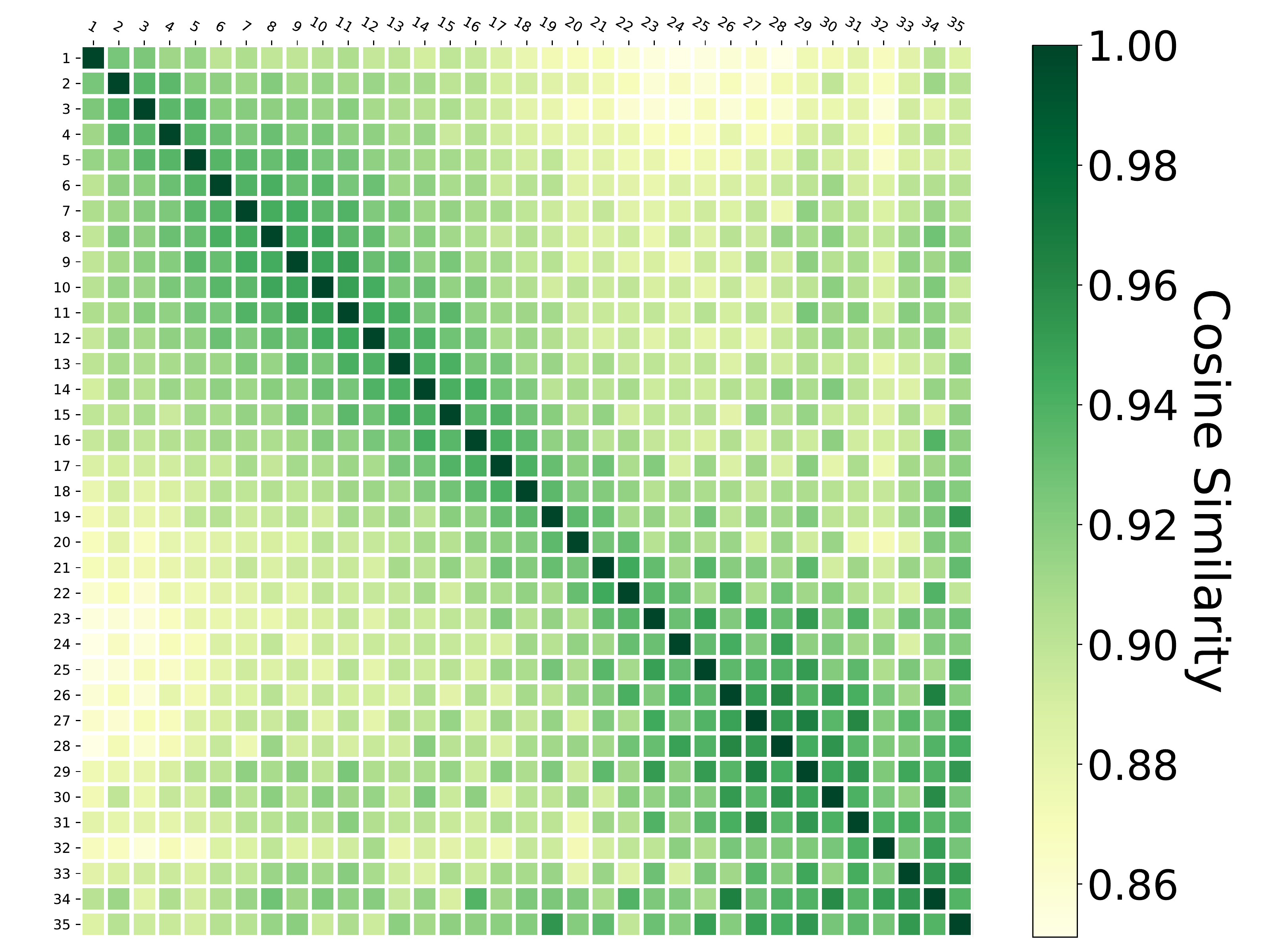}
     \caption{Cosine similarity}
     \label{subfig:cos_feature}
  \end{subfigure}
  \begin{subfigure}[b]{0.49\linewidth}
    \includegraphics[width=\linewidth]{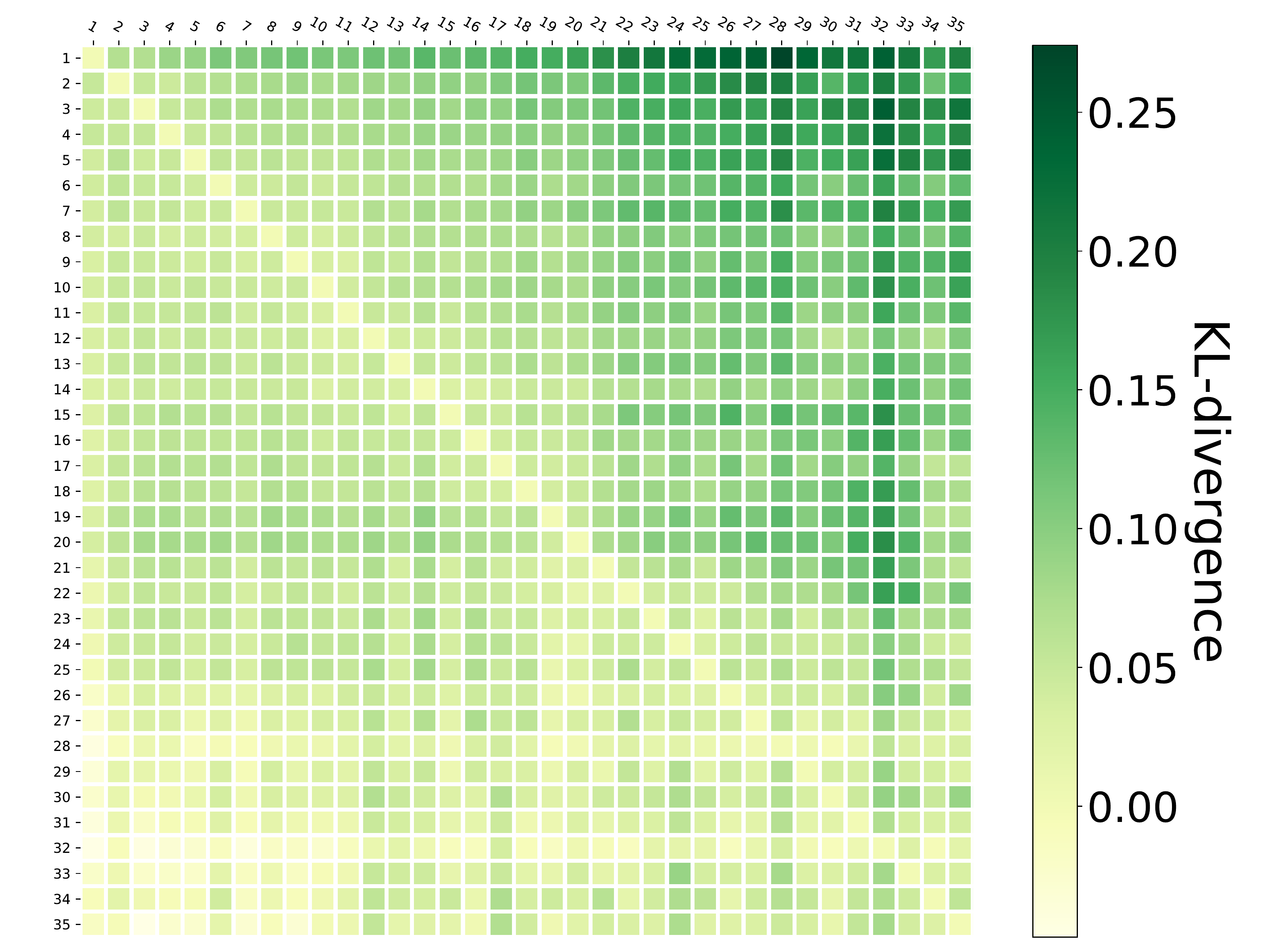}
    \caption{KL divergence}
    \label{subfig:kl_feature}
  \end{subfigure}
\caption{Graph pattern similarity of query sequence $Q$ in sessions.}
\label{fig:kl_cos_feature}
\end{figure}

\noindent\textbf{Graph Pattern Similarity:} Based on~\cite{watdiv2014,saleem2019representative}, we select the following $10$ structural features to form a feature vector\footnote{We append a $1$ to vectors to avoid all-zero vectors.}\footnote{Features in this vector can also be extended to more dimensions or features.}:
\begin{equation}
\label{equ:featureVec}
\begin{split}
        V\ =\ [\ &\#triplePatterns,\ \#BGP,\ \#Projection,\ \#Sink Join Vertex, \\
        &\#Star Join Vertex,\ \#Hybrid Join Vertex,\ \#Path Join Vertex, \\
        & MaxJoinDegree,\ MinJoinDegree,\ MeanJoinDegree\ ].
\end{split}
\end{equation}

For a query sequence $Q=\{q_1, q_2\cdots q_n\}$ in a single search session $s$, we initialize vectors $\{\mathbf{V_{q_1}, V_{q_2}} \cdots \mathbf{V_{q_n}} \}$ according to Equation.~\ref{equ:featureVec}. Then, we normalize every item ($k$ indexes the item) in a vector by 

\begin{equation}
    \mathbf{\hat{V}_{q_i} } (k) = \frac{\mathbf{V_{q_i}} (k)} {\max_{j=1,\cdots,n} \mathbf{V_{q_j}} (k)}.
\end{equation}

We use two metrics to measure graph pattern similarity between two queries: (1) cosine distance, which is a symmetric measurement defined as $Cosine(\mathbf{\hat{V}_{q_1}} , \mathbf{\hat{V}_{q_2}} ) = Cosine( \mathbf{\hat{V}_{q_2}} , \mathbf{\hat{V}_{q_1}} )$ and performed by:
\begin{equation}
    \label{equ:cos}
    Cosine( \mathbf{\hat{V}_{q_1}} , \mathbf{\hat{V}_{q_2}} ) = \frac{\mathbf{\hat{V}_{q_1}}\cdot\mathbf{\hat{V}_{q_2}}}{ \left\Vert \mathbf{\hat{V}_{q_1}} \right\Vert \left\Vert \mathbf{\hat{V}_{q_2}} \right\Vert}
\end{equation}
(2) KL divergence, which is asymmetric, and performed by: 
\begin{equation}
\label{equ:kl}
    D_{KL}( \mathbf{\hat{V}_{q_1}} || \mathbf{\hat{V}_{q_2}} ) = \sum \mathbf{\hat{V}_{q_1}}(k) \log \frac{ \mathbf{\hat{V}_{q_1}} (k)} { \mathbf{\hat{V}_{q_2}} (k) }.
\end{equation}

Cosine similarity between two vectors ranges from $0$ (complete change) to $1$ (constant). Original KL divergence ranges from $0$ (constant) to $+\infty$. The $+\infty$ is caused by zeros in denominators. To prevent this problem, we only calculate $D_{KL}( \mathbf{\hat{V}_{q_1}} || \mathbf{\hat{V}_{q_2}} ) $ when $\mathbf{\hat{V}_{q_1}} (k) \neq 0\ and\ \mathbf{\hat{V}_{q_2}} (k) \neq 0$ here. The minus, zero, and positive result of $D_{KL}(\mathbf{\hat{V}_{q_1}} || \mathbf{\hat{V}_{q_2}} ) $ indicate that distribution of $\mathbf{\hat{V}_{q_2}} $ is more concentrated than, equal to, or more scattered than distribution of $\mathbf{\hat{V}_{q_1}} $, respectively.
Results are presented in Fig.~\ref{fig:kl_cos_feature} as a $m\times m$ matrix $M$, where $m$ is the longest length of sessions. The rows and columns of $M$ are indexed by the query $q_i$ in a session. $M_{ij}$ in $M$ presents the average cosine similarity (or KL divergence) between vectors of $q_i$ and $q_j$ in single search sessions, \ie, $\mathbf{\hat{V}_{q_i}}$ and $\mathbf{\hat{V}_{q_j}}$. 

\begin{figure}[t]
  \centering
  \begin{subfigure}[b]{0.49\linewidth}
    \includegraphics[width=\linewidth]{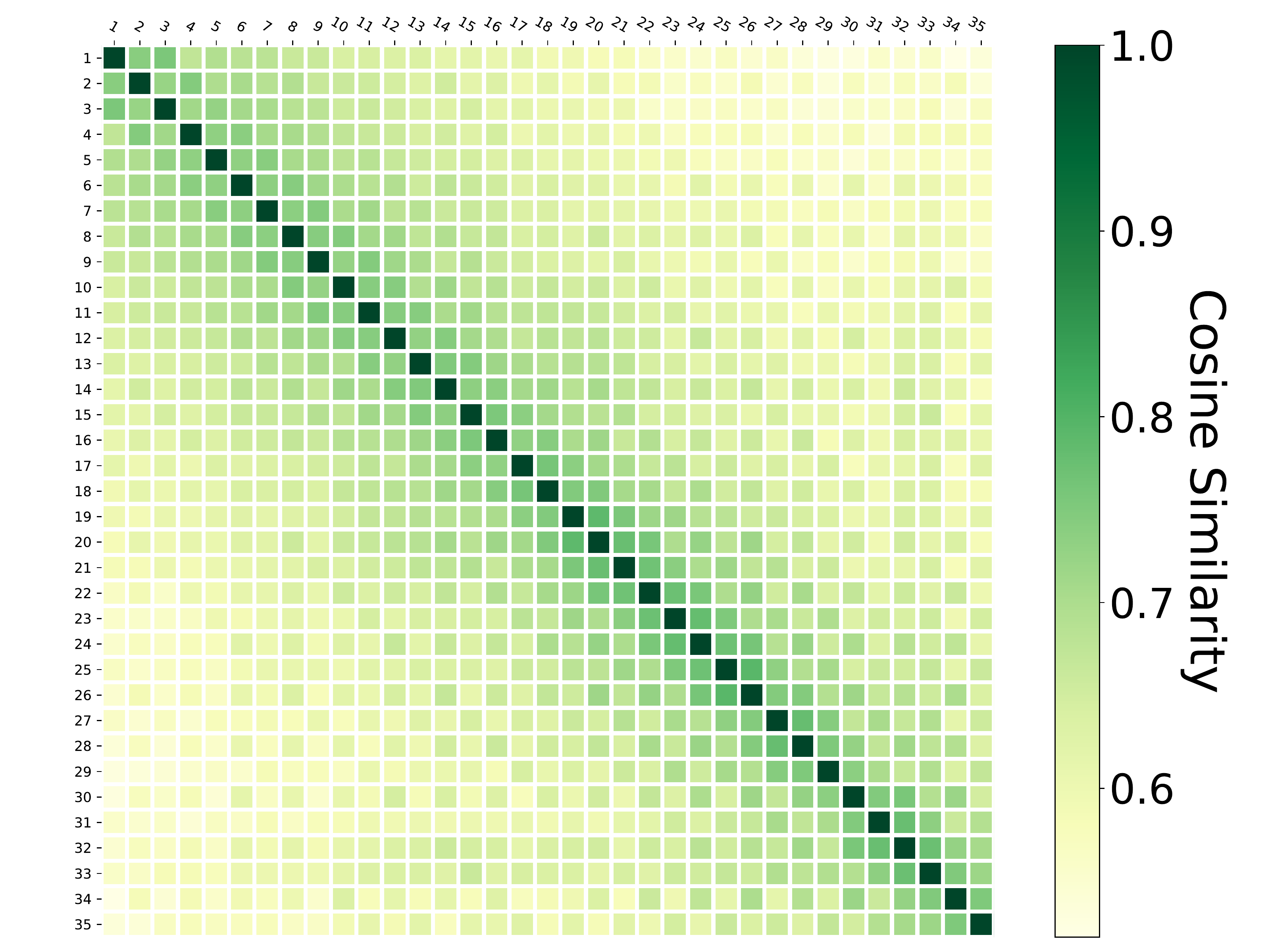}
     \caption{Cosine similarity}
     \label{subfig:cos}
  \end{subfigure}
  \begin{subfigure}[b]{0.49\linewidth}
    \includegraphics[width=\linewidth]{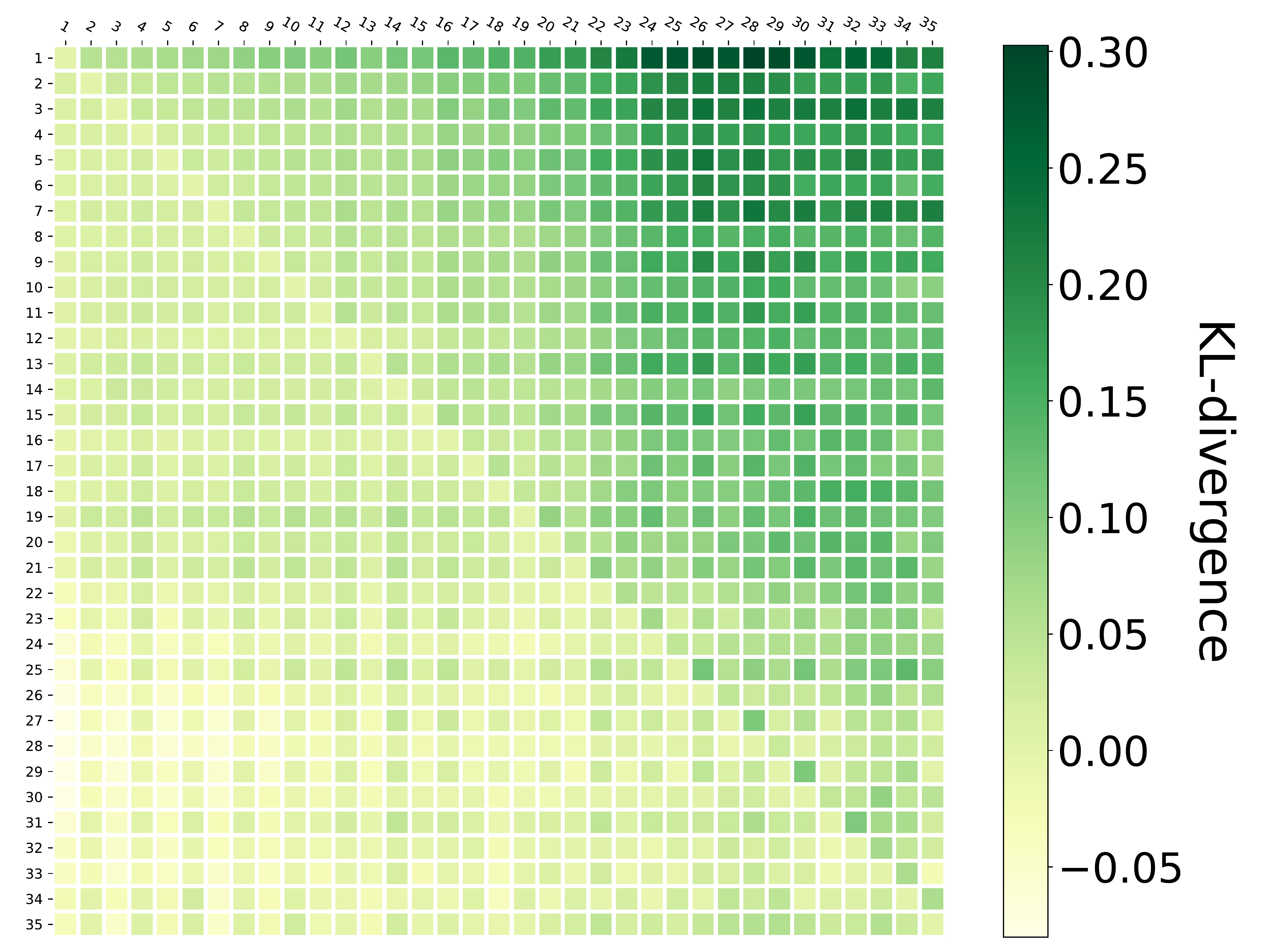}
    \caption{KL divergence}
    \label{subfig:kl}
  \end{subfigure}
\caption{IRI term similarity of query sequence $Q$ in sessions.}
\label{fig:kl_cos}
\end{figure}

\noindent\textbf{IRI Term Similarity:} 
We construct a query-term matrix $D$ which is a $x\times y$ matrix for every dataset. $x$ is the number of queries and $y$ is the number of terms used in queries of a certain dataset. 
The rows of $D$ are indexed by queries $q_i$ and columns are indexed by terms $t_j$. $D_{ij}$ in this matrix represents whether the term $t_j$ is used in query $q_i$. If $t_j$ is used, $D_{ij}$ is $1$. If not, $D_{ij}$ is $0$. 
Query $q_i$ can be represented by a vector $\mathbf{V_{q_i}}$ which is constituted by row $i$ in this query-term matrix $D$. The vector $\mathbf{V_{q_i}}$ indicates the IRI term distribution in $q_i$\footnote{The query representation method can be replaced by other distributed representations such as trained embedding. We do not use embedding here because training embeddings for $10$ datasets is highly resource and time-consuming.}. 
Then, we use cosine similarity and KL divergence to visualize the evolvement of IRI term similarity in single search sessions, as shown in Fig.~\ref{fig:kl_cos}. Please note that the analyses and visualizations of Fig.~\ref{fig:conti_first_graph}, \ref{fig:kl_cos_feature}, \ref{fig:kl_cos} are based on raw data. Lengths of sessions are not normalized. The motivation is to find different user behaviors by session position, which is normally used in log analysis of web searching fields. 

\begin{figure}[t]
  \includegraphics[width=0.9\linewidth]{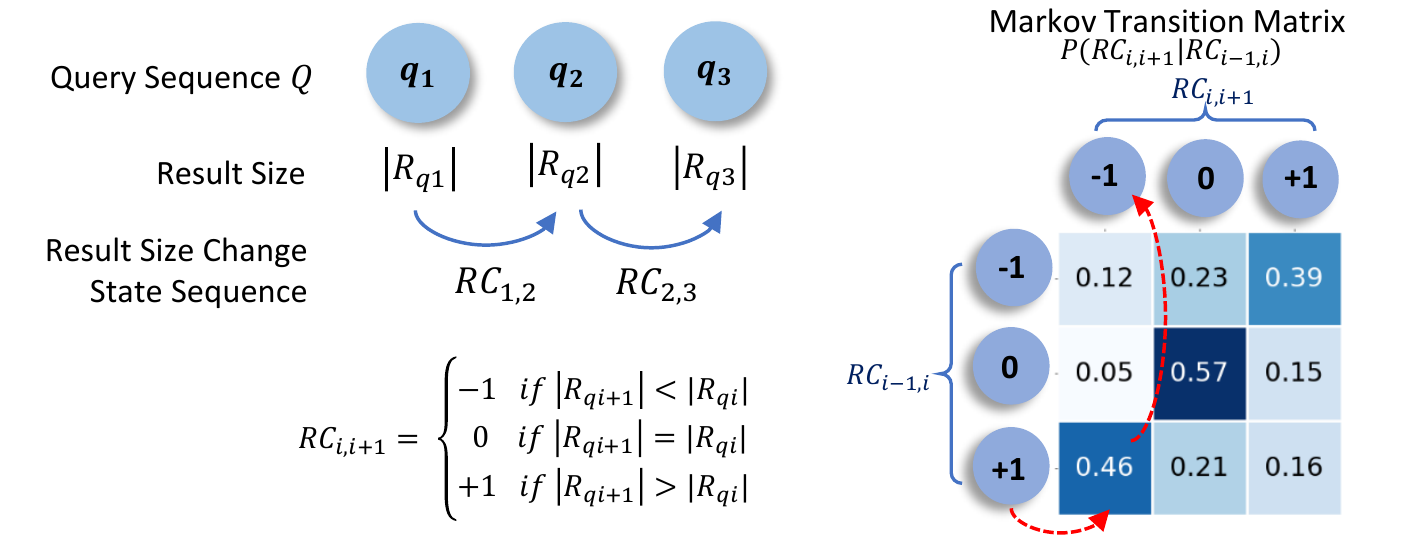}
  \caption{Markov transition matrix of result size change states.}
  \label{fig:trans}
\end{figure}

\noindent\textbf{Change of Result Size:} For the \emph{data-driven} feature, we investigate the transition probability between three result size change states, \ie, decrease, remain unchanged, and increase\footnote{We eliminate the processing error state here.}, presented as $-1$, $0$, $+1$ respectively. For a single query sequence $Q=\{q_1,q_2,\cdots,q_{n-1},q_n\}$, we generate a \emph{result size change state sequence} $\{RC_{1,2},\cdots,RC_{n-1,n}\}$ which is then used to calculate the Markov transition matrix between three types of $RC$ states, \ie, $-1,0,$ and $+1$. A Markov transition matrix is a square matrix describing the probabilities of transferring from one state to another. The state transition probability is formulated by $P(RC_{i,i+1}|RC_{i-1,i}) = \frac{\#(RC_{i-1,i}, RC_{i,i+1})}{\#RC_{i-1,i}}$. 
The state transition probabilities in single search sessions on our $10$ datasets are shown in Fig.~\ref{fig:trans}. The probability of transferring from $RC_{i-1,i}$ to $RC_{i,i+1}$ can be determined by the intersections of the corresponding row and column in the Markov transition matrix. For example, as shown in Fig.~\ref{fig:trans}, the number $0.46$ in row $+1$, column $-1$ indicates the probability of moving from a result size increase state to a decrease state, \ie, $P(-1|+1)$, is $0.46$.


\noindent\textbf{Key Findings:} The above results allow us to make the following observations.
\begin{enumerate}
    \item The GEDs between $(q_i, q_{i+1})$ decrease gradually in a session (as shown in Fig.~\ref{subfig:conti}), which indicates that query change between two contiguous queries is increasingly indistinct as users getting closer to their information needs. Interestingly, the GEDs between $(q_1, q_{i+1})$ increase consistently at first, then decrease (Fig.~\ref{subfig:first}). 
    Combining with key findings.3 (below), this suggests that users may use prior query structures to explore other related information.

    \item The graph patterns of queries in the same session are broadly similar, as illustrated by the $0.78 \sim 1$ cosine similarity and $-0.07 \sim 0.46$ KL divergence in Fig.~\ref{fig:kl_cos_feature}. This indicates that users usually change the structure of graph patterns slightly in a SPARQL search session. The GEDs in Fig.~\ref{fig:conti_first_graph} show this conclusion as well.

    \item The IRI term similarity is less and less similar, as shown in Fig.~\ref{fig:kl_cos}. In more detail, the distribution of IRI terms used in queries is more scattered, which indicates that users tend to include more IRI terms and express a clearer intention as the session moves forward. However, please note that IRI terms are not getting entirely different considering high numerical values (cosine similarity ($0.5 \sim 1.0$) and kl-divergence (up to $0.30$) in Fig.~\ref{fig:kl_cos}).
    
    \item The result size changes indicate that previous result size change does influence the intention of the current query: if the number of results is increased (decreased) in the last query, then users tend to make their current queries more specific (general). On the other hand, the same number of results (mostly zero results) indicates the unfamiliarity of underlying data for users, which may lead to an additional iteration of zero results.
\end{enumerate}

\subsection{Query Reformulations in Single SPARQL Sessions}
We explore different types of reformulations over query sequences in single SPARQL search sessions as discussed in detail below. Please note that we have not considered semantically equivalent rewritings for now. We only track user reformulations and try to find valuable findings about user behaviours.


\noindent\textbf{Reformulations of SPARQL Operators:}
\label{subsec:op} 
We first investigate reformulations in terms of operators for contiguous query pairs. The usage of operators generally reflects the query intent. For instance, the addition of \texttt{Distinct} may occur when a user checks answers with many duplicates.
For SPARQL query forms (i.e., \texttt{SELECT, CONSTRUCT, ASK, DESCRIBE}), only $2.51\%$ SPARQL query pairs have such changes. Among these changes, the most common one ($49.26\%$ relative to the number of changes on query forms) is between \texttt{SELECT} and \texttt{CONSTRUCT}. 
This indicates that users first check the underlying data in knowledge graphs by \texttt{SELECT}, then construct a graph using \texttt{CONSTRUCT}. 
The second most frequent change ($22.60\%$) of the query form is between \texttt{SELECT} and \texttt{ASK}. In this scenario, users may issue an \texttt{ASK} query to examine whether a specific solution exists, followed by the \texttt{SELECT} query to get the desired results. 
Tab.~\ref{tab:op} shows the percentage-wise distribution of the different operators in terms of their usage (i.e., presence of an operator) and reformulations (addition or removal) in the query logs. Please note that \emph{removals} and \emph{additions} are relative to the total \emph{usage} of the operators in the query logs. The results show that a majority of the selected operators are frequently reformulated in the query logs.   

\noindent\textbf{Reformulations of Triple Patterns:}
\label{subsec:changesOftp}
Given a contiguous query pair $(q_i, q_{i+1})$ in a search session, there are three formulations possible pertaining to the triple patterns used in the query pair: (1) new triple pattern(s) is added, (2) existing triple pattern(s) is deleted, (3) some changes (\emph{substitutions}) are made in the individual elements 
(\emph{subject, predicate, object}) of the triple patterns. 
In this section, we show reformulations of triple patterns within different SPARQL operator blocks (e.g., \texttt{Union}, \texttt{Optional}) and the substitutions made on different join vertex types and their connecting edges and nodes.

\begin{table}[t]
\centering
\caption{Percentages ($\%$) of the total usage and reformulations (removals and additions) of operators over all query logs. Coloring is used to show different groups of operators, namely, graph patterns, property paths, aggregations, and solution modifiers.}
\label{tab:op}
\resizebox{\textwidth}{!}{%
\begin{tabular}{lccc|lccc}
\toprule
 Operator & Total-usage & Removals & Additions & Operator & Total-usage & Removals & Additions \\
 \midrule
\rowcolor[HTML]{e6ffe6} 
Filter & 64.67 & 11.13 & 10.99 & \cellcolor[HTML]{ffffcc}Count & \cellcolor[HTML]{ffffcc}2.64 & \cellcolor[HTML]{ffffcc}38.06 & \cellcolor[HTML]{ffffcc}31.00 \\
\rowcolor[HTML]{e6ffe6} 
Union & 23.42 & 9.63 & 9.56 & \cellcolor[HTML]{ffffcc}Sample & \cellcolor[HTML]{ffffcc}0.17 & \cellcolor[HTML]{ffffcc}27.19 & \cellcolor[HTML]{ffffcc}32.46 \\
\rowcolor[HTML]{e6ffe6} 
Optional & 15.14 & 18.62 & 19.02 & \cellcolor[HTML]{ffffcc}GroupConcat & \cellcolor[HTML]{ffffcc}0.10 & \cellcolor[HTML]{ffffcc}23.19 & \cellcolor[HTML]{ffffcc}20.29 \\
\rowcolor[HTML]{e6ffe6} 
Graph & 1.71 & 25.33 & 26.45 & \cellcolor[HTML]{ffffcc}Sum & \cellcolor[HTML]{ffffcc}0.03 & \cellcolor[HTML]{ffffcc}50.00 & \cellcolor[HTML]{ffffcc}50.00 \\
\rowcolor[HTML]{e6ffe6} 
Bind & 0.98 & 24.55 & 25.00 & \cellcolor[HTML]{ffffcc}Min & \cellcolor[HTML]{ffffcc}0.06 & \cellcolor[HTML]{ffffcc}12.50 & \cellcolor[HTML]{ffffcc}15.00 \\
\rowcolor[HTML]{e6ffe6} 
Minus & 0.37 & 52.63 & 42.91 & \cellcolor[HTML]{ffffcc}Max & \cellcolor[HTML]{ffffcc}0.02 & \cellcolor[HTML]{ffffcc}45.45 & \cellcolor[HTML]{ffffcc}27.27 \\
\rowcolor[HTML]{e6ffe6} 
Service & 0.33 & 17.49 & 21.08 & \cellcolor[HTML]{ffffcc}Avg & \cellcolor[HTML]{ffffcc}0.02 & \cellcolor[HTML]{ffffcc}14.29 & \cellcolor[HTML]{ffffcc}28.57 \\
\rowcolor[HTML]{e6ffe6} 
Values & $<$0.01 & 0 & 100 & \cellcolor[HTML]{ffe6ff}Distinct & \cellcolor[HTML]{ffe6ff}50.63 & \cellcolor[HTML]{ffe6ff}4.08 & \cellcolor[HTML]{ffe6ff}3.73 \\
\rowcolor[HTML]{e6e6ff} 
SeqPath & 2.22 & 4.20 & 5.93 & \cellcolor[HTML]{ffe6ff}Limit & \cellcolor[HTML]{ffe6ff}21.84 & \cellcolor[HTML]{ffe6ff}13.58 & \cellcolor[HTML]{ffe6ff}14.00 \\
\rowcolor[HTML]{e6e6ff} 
MulPath & 0.67 & 11.62 & 12.06 & \cellcolor[HTML]{ffe6ff}OrderBy & \cellcolor[HTML]{ffe6ff}7.97 & \cellcolor[HTML]{ffe6ff}25.83 & \cellcolor[HTML]{ffe6ff}24.29 \\
\rowcolor[HTML]{e6e6ff} 
AltPath & 0.30 & 7.43 & 26.73 & \cellcolor[HTML]{ffe6ff}Offset & \cellcolor[HTML]{ffe6ff}2.60 & \cellcolor[HTML]{ffe6ff}16.06 & \cellcolor[HTML]{ffe6ff}21.58 \\
\rowcolor[HTML]{e6e6ff} 
InvPath & 0.05 & 14.29 & 8.57 & \cellcolor[HTML]{ffe6ff}GroupBy & \cellcolor[HTML]{ffe6ff}1.35 & \cellcolor[HTML]{ffe6ff}23.85 & \cellcolor[HTML]{ffe6ff}26.04 \\
\rowcolor[HTML]{ffe6ff} 
Projection & 67.77 & 3.29 & 3.34 & Having & 0.22 & 18.00 & 26.67 \\
\bottomrule
\end{tabular}%
}
\end{table}

\textbf{(1) Operator Block-wise Reformulations:} 
There are $78.24\%$ pairs of contiguous queries that show changes in triple patterns. We list the triple pattern reformulations that occur in the top-$6$ most frequent SPARQL operator blocks in Tab.~\ref{tab:triple_change}. In addition, we show reformulations in the \texttt{Main} block which does not contain any of the operator blocks. An example of the \texttt{Main} block is the body of the SPARQL \texttt{SELECT} query, which only contains a set of triple patterns as the BGP. The triple patterns in the \texttt{Graph Template} operator block represent the graph templates used in the \texttt{CONSTRUCT} queries. The \emph{combined substitutions} represent the made-in-one or more-than-one elements of the triple patterns. The percentages reported in this table are relative to the total usage of certain operator blocks in query logs. 
Note that the reformulations on triple patterns are only considered when the corresponding operator block exists both in $q_i$ and $q_{i+1}$. The additions and removals of operators are not included in Tab~\ref{tab:triple_change}.

The results suggest that reformulations are very common in different operator blocks. 
In most operator blocks, substitutions are more frequent than additions and removals. This indicates that users first make changes in the elements of the existing triple patterns, rather than inserting or deleting new triple patterns.
Substitutions happening in \emph{subject, predicate} and \emph{object} are mostly evenly distributed. 

\begin{table}[t]
\centering
\caption{Percentages ($\%$) of reformulations (additions, removals, and substitutions) on triple patterns in 6 most frequent operator blocks, as well as percentages ($\%$) of substitutions occurring in different elements of triple patterns. (Template = Graph Template).}
\label{tab:triple_change}
\resizebox{\textwidth}{!}{%
\setlength{\tabcolsep}{0.44em}

\begin{tabular}{fgeeeeee}
\toprule
             & Main           & \ Template    & \ \  Union       & Optional       & \ Service     & \ Graph        & \ Subquery       \\
             \midrule
Addition    & 21.92 & 42.00 & 7.79    & 4.30    & 6.87   & 4.21   & 79.95 \\
Removal     & 21.28 & 26.33  & 8.30    & 3.00    & 7.63  & 3.31   & 80.06 \\
Combined Substitution & 61.10 & 55.92 & 43.16 & 17.76 & 9.16  & 8.93  & {\color[HTML]{3531FF}1.21}   \\
\midrule
\emph{Subject} substitution      & 37.97 & 28.95   & 49.81 & 40.41   & 54.55 & 18.66 & 73.88   \\
\emph{Predicate} substitution    & 28.06 & 36.44 & {\color[HTML]{3531FF}18.49} & {\color[HTML]{3531FF}9.56}    & 18.18  & 36.27 & {\color[HTML]{3531FF}5.22}     \\
\emph{Object} substitution       & 33.97 & 34.61 & 31.70 & 50.02 & 27.27  & 45.07 & 20.90    \\
\bottomrule
\end{tabular}%
}
\end{table}

\begin{table}[t]
\centering
\caption{Percentages ($\%$) of substitutions on join vertex types and neighbors. (neigh = neighbor, in = incoming, out = outgoing)}
\label{tab:structure}
\resizebox{\textwidth}{!}{%
\begin{tabular}{ffeeeeeeebad}
\toprule
 &  & Ncbigene & Ndc & Orphanet & Sgd & Sider & Affymetrix & Goa & SWDF & LGD & DBpedia \\
 \midrule
 & center & 20.51 & 12.50 & 11.40 & 1.98 & 20.37 & 2.08 & 1.96 & 6.93 & 28.48 & 29.38 \\
 & neigh edges & 3.20 & 2.37 & 3.35 & 11.44 & 0.77 & 3.75 & 2.88 & 3.77 & 8.63 & 7.11 \\
\multirow{-3}{*}{Star} & neigh nodes & 12.32 & 9.34 & 14.13 & 13.14 & 35.00 & 20.13 & 18.86 & 6.40 & 9.38 & 12.67 \\
\midrule
 &center & 5.26 & 5.88 & 3.70 & 8.68 & {\color[HTML]{FE0000} 45.75} & 2.94 & 0.64 & 7.41 & 0.64 & 15.52 \\
 &neigh edges & 0 & 5.88 & 0 & 4.25 & 0.08 & 22.73 & 0 & 4.00 & 0.27 & 11.96 \\
\multirow{-3}{*}{Sink} & neigh nodes & 19.51 & 10.73 & 22.22 & 51.73 & 11.18 & 65.24 & 61.83 & 6.62 & 26.06 & 22.12 \\
\midrule
 & center & 26.67 & 22.81 & 21.43 & 6.19 & 6.28 & 17.13 & 24.53 & 7.91 & 29.04 & 27.28 \\
\multirow{-2}{*}{Path} & neigh edges & 1.67 & 5.26 & 7.14 & 13.84 & 0.88 & 7.69 & 8.49  & 4.73 & 14.31 & 5.42 \\
\midrule
 & center & 18.75 & 0 & 0 & {\color[HTML]{FE0000} 90.88} & 13.04 & {\color[HTML]{FE0000} 87.53} & {\color[HTML]{FE0000} 68.84} & {\color[HTML]{FE0000} 11.51} & {\color[HTML]{FE0000} 43.92} & {\color[HTML]{FE0000} 45.54} \\
 & in edges & 0 & 0 & 0 & 0 & 0 & 0 & 0 & 0.66 & 0.80 & 15.19 \\
 & in nodes & 0 & 0 & 0 & 37.15 & 2.94 & 36.06 & 28.70 & 5.81 & 18.73 & 24.41 \\
 & out edges & 23.81 & 19.23 & 30.00 & 0.38 & 0 & 0.18 & 1.27 & 1.12 & 0.27 & 5.72 \\
\multirow{-5}{*}{Hybrid} & out nodes & {\color[HTML]{FE0000} 47.62} & {\color[HTML]{FE0000} 41.67} & {\color[HTML]{FE0000} 60.00} & 2.62 & 16.83 & 1.68 & 7.84 & 4.45 & 18.31 & 13.66 \\
\bottomrule
\end{tabular}%
}
\end{table}

\textbf{(2) Substitutions on Different Join Vertex Types and Neighbors:} 
To further study user preferences on substitutions, we investigate the elements in hypergraphs of the queries in which substitutions appear most frequently. To this end, we consider the join vertex (the center), the direct \emph{subjects} and \emph{objects} this vertex connects to (neighbor nodes), and the direct \emph{predicates} this vertex connects to (neighbor edges). Also, for \emph{hybrid} vertices, we divide their neighbors into incoming and outgoing types.
The distribution of substitutions on different positions is shown in Tab.~\ref{tab:structure}.
These percentages are relative to the occurrence of join vertex types and the neighbors of centers. We use \emph{red} to indicate the highest value in each column. 
The results indicate that most substitutions happen on incoming nodes of the \emph{hybrid} vertex. This is because the hybrid node has the highest connectivity with other nodes in the query hypergraph. As such, it is likely to be changed more frequently by the users. 
\begin{figure}[t]
  \includegraphics[width=\linewidth]{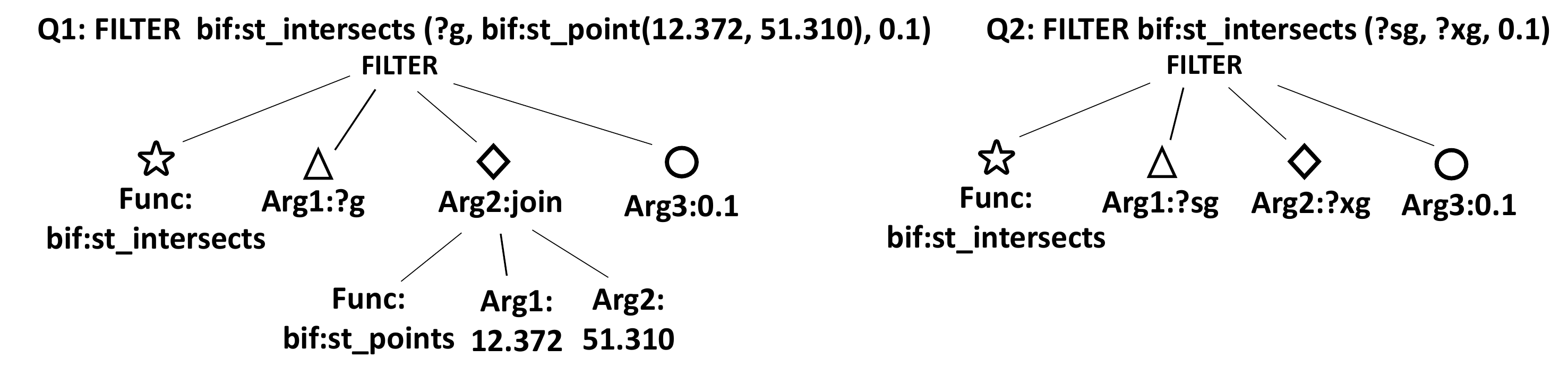}
  \caption{Example of corresponding elements of two \texttt{FILTER} constraints.}
  \label{fig:filterSample}
\end{figure}


\begin{table}[t]
\centering
\caption{Substitutions on \texttt{FILTER} constraints.}
\label{tab:filter_change}
\resizebox{\textwidth}{!}{%
\begin{tabular}{feeeeeeebad}
\toprule
Dataset          & Ncbigene & Ndc     & Orphanet & Sgd     & Sider      & Affymetrix & Goa  & SWDF   & LGD     & DBpedia \\
\midrule
\emph{Block Subs}    & 0        & 20      & 8        & 121     & 3      & 94         & 7    & 34   & 11,787  & 3,155   \\
\emph{Specific Subs} & 4        & 69      & 52       & 67      & 32     & 97         & 17  & 137    & 5,017   & 5,702   \\
\midrule
variable       & 25.00  & 24.64 & 21.15  & 17.91 & 0    & {\color[HTML]{FE0000}39.18}    & 11.76  & 19.71 & 3.41  & 21.43 \\
IRI              & 0   & {\color[HTML]{FE0000}59.42} & 5.77   & {\color[HTML]{FE0000}74.63} & 0   & 20.62    & 35.29 & 8.03 & 1.14  & {\color[HTML]{FE0000}30.90} \\
string           & {\color[HTML]{FE0000}75.00}  & 15.94 & {\color[HTML]{FE0000}73.08}  & 7.46  & {\color[HTML]{FE0000}100}  & 38.14    & 29.41 & {\color[HTML]{FE0000}64.96} & 3.37  & 28.20 \\
number           & 0   & 0  & 0   & 0  & 0  & 0     & 0  & 0.73  & {\color[HTML]{FE0000}90.53} & 7.33  \\
\bottomrule
\end{tabular}%
}
\end{table}

\noindent\textbf{Reformulations of FILTER Constraints: }
A constraint, expressed by the keyword \texttt{FILTER}, is a restriction on solutions over the whole group in which the \texttt{FILTER} appears\footnote{\url{https://www.w3.org/TR/sparql11-query/#scopeFilters}}. Similar to reformulations of triple patterns, there are three reformulations for \texttt{FILTER} constraints: addition(s), removal(s), and substitutions on elements of \texttt{FILTER} constraints. But unlike the triple pattern, which has three elements, a \texttt{FILTER} constraint can have different elements that are used inside its body. Therefore, we express \texttt{FILTER} constraints as parse trees (see~Fig.\ref{fig:filterSample}). On this basis, we compare two parse trees of \texttt{FILTER} constraints in contiguous queries, \ie, $q_i$ and $q_{i+1}$, and find changes in corresponding elements (\emph{substitutions}). For example, the elements with the same patterns in Fig.~\ref{fig:filterSample} are compared.
Please note that corresponding elements are not only between single values, \ie, \emph{specific substitutions}, which can be compared directly, but also between one single value and another constraint function or between constraint functions \ie, \emph{block substitutions}. The \emph{Arg2} in Fig.~\ref{fig:filterSample} is an example of \emph{block substitution}.

Among all the contiguous query pairs in the same session, $37.68\%$ have reformulations about \texttt{FILTER} constraints. Again, the majority of the reformulations ($92.17\%$) are substitutions. 
We present distributions of \emph{block} and \emph{specific substitutions} in different datasets in Tab.~\ref{tab:filter_change}. The first two rows present the number of \emph{block} and \emph{specific substitutions}. 
For \emph{specific substitutions}, we also list the percentage of \emph{specific substitutions} on different data types: variable, IRI, string, and number. 
Substitutions on \texttt{FILTER} constraints show very different distributions in different datasets, especially in the LGD. The LGD is a knowledge graph that contains spatial datasets~\cite{linkedgeodata}. Geometry data types and functions embedded in GEOSPARQL~\cite{geosparql} are used to satisfy the needs for representing and retrieving spatially related data. 
Functions like \emph{intersects, overlaps} involve many numeric calculations, which makes \emph{numbers-specific substitutions} more dominant ($90.53\%$) compared to other substitutions. Furthermore, \texttt{FILTER} constraints in the LGD are more complex than other datasets: the number of \emph{block substitutions} is twice the number of \emph{specific substitutions}. For other datasets, substitutions usually happen in strings, which serve as an argument in functions of strings. 


\begin{figure}[t]
  \includegraphics[width=\linewidth]{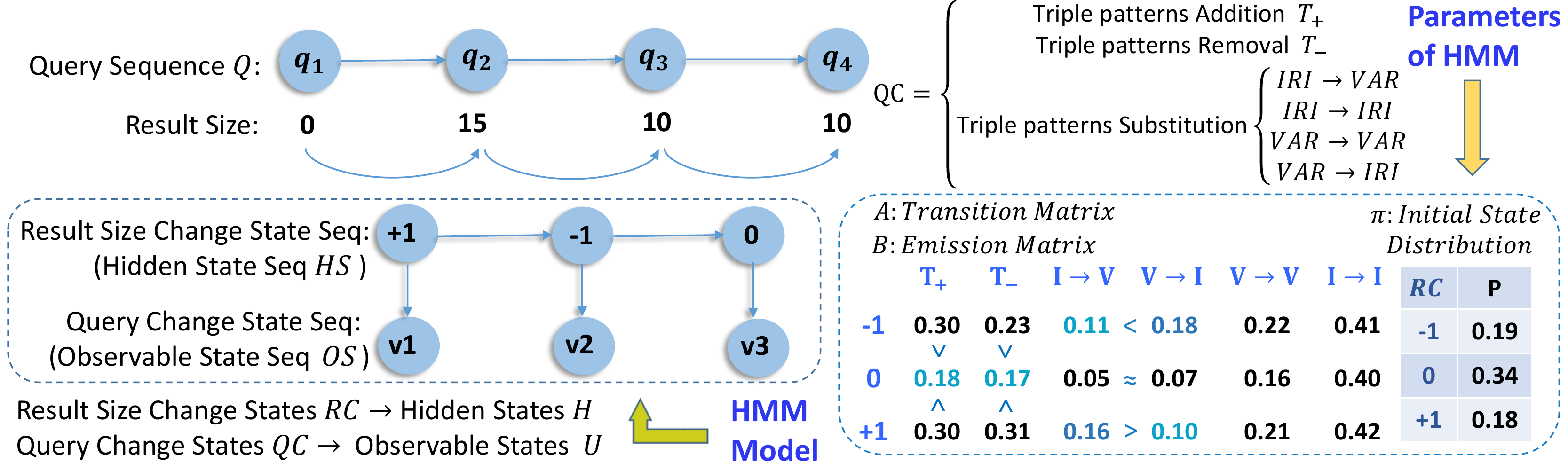}
  \caption{HMM model of the SPARQL search session.}
 \label{fig:hmm}
\end{figure}


\section{An Application Example of Findings}
\label{sec:demo}
To show the potentiality of our findings, we use a simple model to predict user intentions of a given session and give reformulation suggestions based on the predicted intention.
%
%
In single sessions, continuous query reformulations can be captured, while the drifting user intentions are abstract and can not be observed directly. 
We utilize a Hidden Markov Model (HMM) to characterize this process, and model the intention sequence as an unobservable sequence, the reformulation sequence as an observable sequence.
%
We assume that $H$=$\{h_1, h_2 \cdots h_e\}$ is the set of hidden states and $U$=$\{u_1, u_2 \cdots u_f\}$ is the set of observable states ($e$ and $f$ are the number of corresponding states), while $HS$=$(hs_1, hs_2\cdots hs_t)$ and $OS$=$(os_1, os_2\cdots os_t)$ are the sequence of hidden states and observable states, respectively. We use result size changes to model drifting intentions, \ie, the hidden states $H$, and consider changes in triple patterns as observable states $U$. 
We employ the maximum likelihood estimation to calculate parameters of the HMM model, \ie, initial state distribution $\pi$, transition probability matrix $A$ for hidden states, and the emission probability matrix $B$.
The matrix $A$ is the Markov transition matrix in Fig.~\ref{fig:trans}.
The details and parameters of the model are presented in Fig.~\ref{fig:hmm}.
In summary, our model is capable of (1) inferring user intentions based on the observable query change sequence, \ie, a decoding problem: given $\lambda$=$(A,B,\pi)$ and $OS$=$(os_1, os_2 \cdots os_t)$, calculate a sequence $HS$=$(hs_1, hs_2 \cdots hs_t)$ that maximizes $P(HS|OS)$; (2) suggesting reformulation strategies by the possibility of subsequent reformulation strategies, \ie, an evaluation problem: given parameters of HMM $\lambda$=$(A,B,\pi)$, calculate $p(OS|\lambda)$.

\begin{figure}[t]
\centering
  \includegraphics[width=\linewidth]{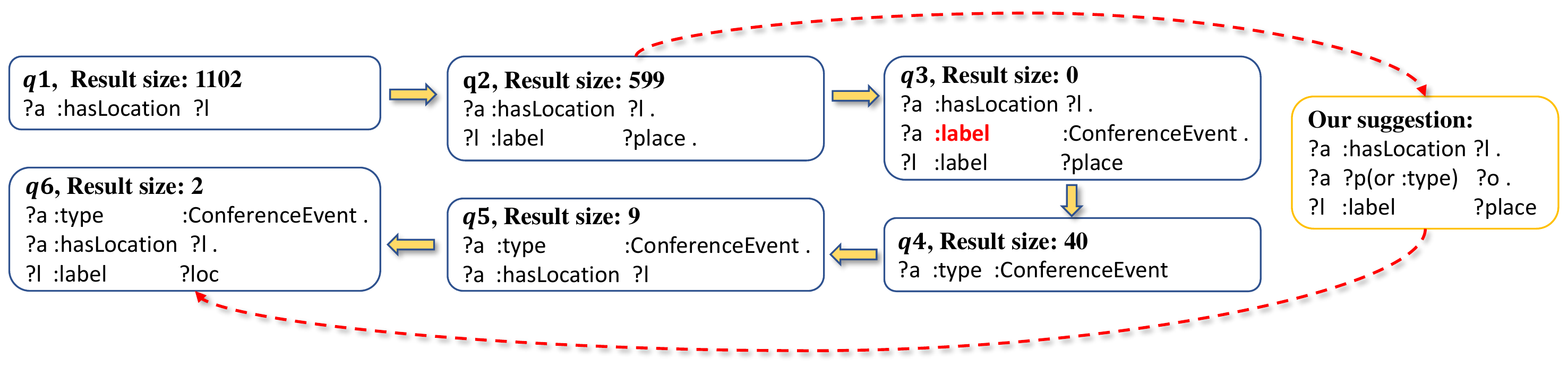}
  \caption{A session (only BGPs without prefixes) from SWDF log dataset.}
  \label{fig:case}
\end{figure}

\noindent\textbf{Case Study:} We use a session shown in Fig.~\ref{fig:case} to illustrate how our model works. 
In this session, the user use $6$ queries in total to find the locations of the \emph{conferenceEvent}. By recommending adding a generic triple pattern to impose a restriction on the variable $a$ first, then making it specific according to the retrieved results, we assist this user omitting a few inefficient reformulations.

\noindent\textbf{Discussions: } %
Here are some possible directions to update the model: (1) More comprehensive ways to model states in HMM. (2) Consider correlations between queries in query sequences, not just correlations of contiguous query pairs. (3) Combination with techniques based on the underlying data.

\section{Related Work}
\label{sec:related}

\noindent \textbf{SPARQL Query Log Analysis in Session Level: } 
Previous SPARQL query log analyses have given a comprehensive analysis of SPARQL queries in terms of \emph{structural} features in isolation, such as the most frequent patterns~\cite{2017pattern}, topological structures~\cite{bonifati2019analytical} and the join vertex types~\cite{2011empirical} \etc. However, the analysis of potential correlations between queries is limited.
Raghuveer~\etal~\cite{2012MachineLoopPattern} first define the query sequence executed by the same user as a \emph{SPARQL user session}. On this basis, they introduce the important feature of robotic queries, the loop patterns, to describe the repetitive patterns of sessions. Bonifati~\etal~\cite{bonifati2019analytical} illustrate the similarity of queries in a query sequence as per the distribution of the length of so-called \emph{streaks}. 
However, the similarity property introduced in these studies~\cite{2012MachineLoopPattern,bonifati2019analytical} applies more to robotic queries from which organic queries are not separated. 
Bonifati~\etal~\cite{2019navigatingwiki} analyze robotic and organic queries separately, but not at the session level.
In this paper, we seek a more specific definition of the \emph{search session}, and comprehensively analyze the evolvements of \emph{structure} and \emph{data-driven} features of the organic queries in sessions.

\noindent \textbf{Applications based on Session-Level Analysis: }
The similarity property proposed by previous researchers~\cite{2012MachineLoopPattern,bonifati2019analytical} has been utilized in query augmentation~\cite{2013detectSplTemp} to retrieve more related results. 
Lorey~\etal~\cite{2013detectSplTemp} focus on retrieving SPARQL queries with high similarity with current queries, but fail to capture the drifting intentions behind queries in single search sessions.
Utilizing explicit user feedback directly, Lehmann~\etal~\cite{autoSPARQL} use active learning to determine different reformulation strategies, which is similar to the example we conduct in this study. However, explicit feedback is usually unavailable in most search scenarios. 
We utilize reformulations between queries as implicit user feedback.

\noindent \textbf{SPARQL Similarity: }
Different SPARQL similarity measures have been used in this paper to capture the query changes in sessions.
Dividino~\etal~\cite{2013SumSimMeasures} classify SPARQL similarity measures into $4$ categories: (1) query structure, where SPARQL queries are expressed as strings or graph structures; (2) query content, where triple patterns are expressed as strings, ontological terms, or numerical values; (3) query languages, based on operators usages; (4) result sets. Dividino~\etal~\cite{2013SumSimMeasures} also argue that the choice of different measurements is dependent on the application at hand. In our work, we consider all these dimensions separately and conduct a comprehensive analysis of query changes in the sessions.

\section{Conclusions}
\label{sec:conclusion}
This paper reveals secrets of user search behaviors in SPARQL search sessions by investigating $10$ real-world SPARQL query logs. Specifically, we analyze the evolvement of query changes, w.r.t. \emph{structural} and \emph{data-driven} features of SPARQL queries in sessions, and reach a series of novel findings. We thoroughly investigate reformulations in terms of SPARQL operators, triple patterns, and \texttt{FILTER} constraints. Furthermore, we provide an application example about the usage of our findings. We hope results presented here will serve as a basis for future SPARQL caching, auto-completion, suggestion, and relaxation techniques.

\section*{Acknowledgements}
This work has been supported by by National Natural Science Foundation of China with Grant Nos. 61906037 and U1736204; the German Federal Ministry for Economic Affairs and Energy (BMWi) within the project RAKI under the grant no 01MD19012D, by the German Federal Ministry of Education and Research (BMBF) within the project DAIKIRI under the grant no 01IS19085B, and by the EU H2020 Marie Skłodowska-Curie project KnowGraphs under the grant agreement no 860801; National Key Research and Development Program of China with Grant Nos. 2018YFC0830201 and 2017YFB1002801; the Fundamental Research Funds for the Central Universities.

%
%
%

{\small
\bibliographystyle{splncs04}
\bibliography{sparql}
}
\end{document}